\documentclass[reqno,a4paper,11pt]{amsart}

\usepackage{amsmath,amstext,amsfonts,amsbsy,eucal,amssymb}
\usepackage[latin1]{inputenc}

\numberwithin{equation}{section}

\newtheorem{theorem}{Theorem}[section]
\newtheorem{lemma}[theorem]{Lemma}

\theoremstyle{definition}

\newtheorem{remark}[theorem]{Remark}

\begin{document}

\parskip 4pt
\baselineskip 16pt


\title[On solutions for some class of integrable difference equations]
{On solutions for some class of integrable difference equations}

\author[Andrei K. Svinin]{Andrei K. Svinin}
\address{Andrei K. Svinin, 
Matrosov Institute for System Dynamics and Control Theory of 
Siberian Branch of Russian Academy of Sciences,
P.O. Box 292, 664033 Irkutsk, Russia}
\email{svinin@icc.ru}

%
%

\date{\today}



\begin{abstract}
In this paper we  show that an arbitrary solution of one ordinary difference equation is also a solution for a hierarchy of integrable difference equations. We also provide an example of such a solution that is related to sequence generated by a second-order linear recursion with 2-periodic coefficients.  
\end{abstract}

\maketitle

\section{Introduction}

In this paper we consider an infinite one-parameter  class of    ordinary difference  equations 
\begin{equation}
y_{n+1}\left(\sum_{j=0}^{s-1}y_{n+j}-H^{(s)}\right)=y_{n+s}\left(\sum_{j=0}^{s-1}y_{n+j+2}-H^{(s)}\right)\;\; s\geq 2.
\label{1555}
\end{equation}
Notice that each of  equations (\ref{1555}) contains one arbitrary parameter $H^{(s)}$ and can be rewritten in the form
\[
y_{n+s+1}=R(y_n,\ldots, y_{n+s}; H^{(s)}),
\]
where $R$ is corresponding rational function of its arguments. Equation (\ref{1555}), for some fixed value of $H^{(s)}$, yields a map $\mathbb{R}^{s+1}\rightarrow \mathbb{R}^{s+1}$ for real-valued initial
data $\{y_n : n=1,\ldots, s\}$. By analogy with ordinary differential equations, a function $J=J_n=J(y_n,\ldots, y_{n+s})$ is called a first integral for
difference equation (\ref{1555}) if by virtue of this equation one has $J_{n+1}=J_n$. There are ordinary difference equations which have some special properties which yield a regular behavior of
their solutions. One  such property is  Liouville integrability \cite{Bruschi}, \cite{Veselov}.

Equations of the form (\ref{1555}) appeared in \cite{Svinin1}. They  are interesting in that they share the property of having a Lax pair that indicates their possible Liouville integrability.    
In \cite{Svinin2}, we have shown a method for constructing a set of the first integrals for these equations and also for a wider class of integrable difference equation regardless of their Lax representation. Unfortunately, the property of Liouville integrability for these equations has not been proved  in our works.  We say that these equations are integrable only in the sense that they admit a nontrivial Lax representation that yields a number of the first integrals.

Also note that, as was shown in \cite{Svinin1} that, assuming that $y_n$'s depend on evolutionary parameters $(t_1, t_2,\ldots)$,  equations (\ref{1555}) play the role of compatible constraints for the Volterra  lattice 
\[
\frac{\partial y_n} {\partial t_1}=y_n\left(y_{n+1}-y_{n-1}\right) 
\]
and its hierarchy of  generalized symmetries that can be written in explicit form as \cite{Svinin1}
\begin{equation}
\frac{\partial y_n} {\partial t_s}=y_n\left(S^s_s(n-s+2)-S^s_s(n-s)\right)\;\; s\geq 2. 
\label{67531}
\end{equation}
Here discrete polynomials $S_s^k(n)$ are explicitly defined  by 
\begin{equation}
S_s^k(n)=\sum_{0\leq q_1\leq \cdots \leq q_k\leq s-1} y_{n+q_1+k-1}\cdots y_{n+q_k}.
\label{6753}
\end{equation}
It should be noted that the first integrals constructed in \cite{Svinin2} are described in terms of these discrete polynomials.

In this article we do not discuss  Liouville integrability for equations  (\ref{1555}). Instead, we show a compatibility  of these equations, which also, to a certain extent, reflects the integrability property. In general, it should be said that the integrability property for equations of various nature is often formulated as a compatibility condition of some number of equations. An example is the pairwise compatibility of evolutionary equations (\ref{67531}).

Specifically, we prove the fact that an arbitrary solution of equation (\ref{1555}) corresponding to the simplest case $s=2$, that is,
\begin{equation}
y_{n+1}\left(y_n+y_{n+1}-H^{(2)}\right)=y_{n+2}\left(y_{n+2}+y_{n+3}-H^{(2)}\right)
\label{109}
\end{equation}
is also a solution of equation (\ref{1555})  for all  $s\geq 3$. In these circumstances, it turns out that parameters $H^{(s)}$ are related to each other  by rather nontrivial recurrence relations  which also include two  first integrals $G^{(s)}$ and $G_1^{(s)}$ that will be defined in the following section. Ultimately each $H^{(s)}$, for $s\geq 3$, is unambiguously expressed as a rational function of variables $(y_0, y_1, y_2, H^{(2)})$.

Also in the paper we construct some class of  solutions for (\ref{1555}) that  are yielded by a second-order linear recursion with 2-periodic coefficients
\begin{equation}
T_{n}=P_nT_{n-1}+QT_{n-2},\;\; P_{n+2}=P_n
\label{45321}
\end{equation}
with arbitrary initial conditions. Rather important circumstance here is that recursion (\ref{45321}) defines a great number of interesting integer sequences with applications in various areas of mathematics. 

The paper is organized as follows. In the following section, using the two integrals mentioned above and some substitutions, we show some difference equations closely related to (\ref{1555})  which are of interest in themselves. In particular, \cite{Hone2} we show that equations (\ref{1555}) are related to bilinear difference equations with 2-periodic coefficients. It should be noted that presentation of material in  Section \ref{555} partially  follows the lines of the paper \cite{Hone2}. At the end of Section \ref{555}, we formulate a theorem concerning the above-mentioned compatibility of equations  (\ref{1555}). In Section \ref{888}, we show how the sequences defined by recursion (\ref{45321}) yield  solutions for equation (\ref{109}) and, therefore, for all equations of the form (\ref{1555}).

\section{Difference equations}  \label{555}

To begin with, note that equation (\ref{1555})  has the first integral 
\begin{equation}
G^{(s)}=\left(\sum_{j=0}^s y_{n+j}-H^{(s)}\right) \prod_{j=1}^{s-1}y_{n+j}.
\label{4}
\end{equation}
We can consider this relation as a difference equation of order $s$ with two parameters $G^{(s)}$ and $H^{(s)}$, that is  equivalent to (\ref{1555}). Indeed, it is easy to see that the relation $G^{(s)}(n+1)=G^{(s)}(n)$ is nothing but (\ref{1555}). 
\begin{remark}
The equation (\ref{4}) known as the generalized DTKQ equation was the object of study in the work \cite{Hone2}. It was shown there that for even $s$  the equation (\ref{4}) are related to the $(s, -1)$ periodic reduction of Hirota's discrete KdV equation, that turned out to be Liouville integrable  \cite{Hone}. In turn, for odd $s$,  equations (\ref{4}) are connected to reductions of a discrete Toda lattice \cite{Hone11}. 
It should be also  noticed, that this equation with $H^{(s)}=0$ has appeared in \cite{Demskoi}.  

More general than (\ref{4}) is a two-parameter class of integrable equations
\[
G^{(k, s)}=\sum_{j=0}^k(-1)^jH_j^{(k, s)}S^{k-j}_{s-j+1}(n+j) \prod_{j=k}^{s-1} y_{n+j},\;\; k\geq 1,\;\; s\geq k+1, 
\]
where $H_0^{(k, s)}=1$,  appeared in \cite{Svinin1} and \cite{Svinin2}. Here $(k, s)$-th difference equation involves $k+1$ parameters $\left(G^{(k, s)}, H_1^{(k, s)},\ldots, H_k^{(k, s)}\right)$.
\end{remark}

Let us consider  the first integral \cite{Hone2}, \cite{Svinin2}
\[
G_1^{(s)}=\prod_{j=0}^s y_{n+j} +\sum_{j=1}^{s-1}y_{n+j} \left(\sum_{j=0}^s y_{n+j}-H^{(s)}\right) \prod_{j=1}^{s-1}y_{n+j},
\]
which further will be useful for us. 
In what follows, we use the fact, that  $G_1^{(s)}$ can be rewritten as
\begin{equation}
G_1^{(s)}=\sum_{j=0}^{s-1} y_{n+j}\left(G^{(s)}-\prod_{j=0}^{s-1} y_{n+j}\right)+H^{(s)} \prod_{j=0}^{s-1} y_{n+j}.
\label{4453}
\end{equation}

With the substitution $y_n=w_nw_{n+1}$ equation  (\ref{4}) becomes 
\begin{equation}
\sum_{j=0}^s w_{n+j}w_{n+j+1}=\frac{G^{(s)}}{\prod_{j=1}^{s-1} w_{n+j}w_{n+j+1}}+H^{(s)}.
\label{5}
\end{equation}
By direct calculation, it can be checked that this equation has  $2$-integral \cite{Hone2}
\begin{equation}
\alpha_n^{(s)}=\prod_{j=0}^sw_{n+j}-\frac{G^{(s)}}{\prod_{j=1}^{s-1}w_{n+j}}.
\label{6}
\end{equation}
Indeed, by (\ref{5}), we have 
\begin{equation}
\alpha_{n+1}^{(s)}=\prod_{j=1}^{s-1}w_{n+j}\left(H^{(s)}-\sum_{j=0}^{s-1}w_{n+j}w_{n+j+1}\right)\;\;\mbox{and}\;\; \alpha_{n+2}^{(s)}=\alpha_n^{(s)}.
\label{61}
\end{equation}
\begin{lemma}
The relation
\begin{equation}
\alpha_{n}^{(s)}\alpha_{n+1}^{(s)}=G_1^{(s)}-H^{(s)}G^{(s)}
\label{8798}
\end{equation}
is valid.
\end{lemma}
\textbf{Proof.} By (\ref{6}) and (\ref{61}), we have
\begin{eqnarray}
\alpha_n^{(s)}\alpha_{n+1}^{(s)}&=&\left(\prod_{j=0}^sw_{n+j}-\frac{G^{(s)}}{\prod_{j=1}^{s-1}w_{n+j}}\right)\prod_{j=1}^{s-1}w_{n+j}\left(H^{(s)}-\sum_{j=0}^{s-1}w_{n+j}w_{n+j+1}\right) \nonumber\\
&=&\left(\prod_{j=0}^{s-1} w_{n+j}w_{n+j+1}-G^{(s)}\right)\left(H^{(s)}-\sum_{j=0}^{s-1}w_{n+j}w_{n+j+1}\right). \label{776545}
\end{eqnarray}
On the other hand, with the substitution $y_n=w_nw_{n+1}$,  (\ref{4453}) becomes 
\[
G_1^{(s)}=\sum_{j=0}^{s-1} w_{n+j}w_{n+j+1}\left(G^{(s)}-\prod_{j=0}^{s-1} w_{n+j}w_{n+j+1}\right)+H^{(s)} \prod_{j=0}^{s-1} w_{n+j}w_{n+j+1}.
\]
Comparing the latter with (\ref{776545}), in a result, we get relation (\ref{8798}).
$\Box$

Let us now rewrite (\ref{6}) as 
\begin{equation}
\prod_{j=0}^{s-1} w_{n+j}w_{n+j+1}=\alpha_n^{(s)} \prod_{j=1}^{s-1} w_{n+j}+G^{(s)},\;\; \alpha_{n+2}^{(s)}=\alpha_n^{(s)}.
\label{7}
\end{equation}
This is an ordinary difference equation of order $s$. This equation is known as the modified generalized DTKQ equation \cite{Hone2}. In particular case $s=2$, it becomes
\begin{equation}
w_{n}w_{n+1}^2w_{n+2}=\alpha_n^{(2)} w_{n+1}+G^{(2)},\;\; \alpha^{(2)}_{n+2}=\alpha^{(2)}_n. 
\label{10}
\end{equation}
Equation (\ref{7}) is of interest, because, as was pointed out in \cite{Hone2}, it represents an example of $U$-system \cite{Hone1}.

Let us notice that here $H^{(s)}$ plays the role of the first integral for difference equation (\ref{7}). 
From (\ref{61}), we derive the following expression for this integral:
\begin{equation}
H^{(s)}=\sum_{j=0}^{s-1} w_{n+j}w_{n+j+1}+\frac{\alpha_{n+1}^{(s)}}{\prod_{j=1}^{s-1} w_{n+j}}.
\label{761}
\end{equation}
By direct calculation, it can be checked that,  by (\ref{7}), the first integral (\ref{761}) can be rewritten as
\begin{eqnarray}
H^{(s)}&=&\sum_{j=0}^{s-2} w_{n+j}w_{n+j+1}+\frac{\alpha_n^{(s)}}{\prod_{j=0}^{s-2} w_{n+j}} +\frac{\alpha_{n+1}^{(s)}}{\prod_{j=1}^{s-1} w_{n+j}} \nonumber\\
&&+\frac{G^{(s)}}{\prod_{j=0}^{s-2} w_{n+j}w_{n+j+1}}. \label{765}
\end{eqnarray}

In particular, in the case $s=2$, (\ref{765}) becomes
\begin{equation}
H^{(2)}=w_{n}w_{n+1}+\frac{\alpha_{n}^{(2)}}{w_{n}}+\frac{\alpha_{n+1}^{(2)}}{w_{n+1}}+\frac{G^{(2)}}{w_{n}w_{n+1}}.
\label{9809}
\end{equation}

With the substitution $w_n=f_nf_{n+2}/f_{n+1}^2$ equation (\ref{7}) can be rewritten in bilinear form \cite{Hone2}
\begin{equation}
f_nf_{n+s+2}=\alpha_n^{(s)} f_{n+1}f_{n+s+1}+G^{(s)} f_{n+2}f_{n+s},\;\; \alpha^{(s)}_{n+2}=\alpha^{(s)}_n.
\label{8}
\end{equation}
More exactly, any solution of (\ref{8}) yields a solution of (\ref{7}) but not vice versa. In particular, if $s=2$, then (\ref{8}) becomes
\begin{equation}
f_nf_{n+4}=\alpha_n^{(2)} f_{n+1}f_{n+3}+G^{(2)} f_{n+2}^2,\;\; \alpha^{(2)}_{n+2}=\alpha^{(2)}_n.
\label{9}
\end{equation}

\begin{remark}
In fairness, we note that the substitution $y_n=f_nf_{n+3}/(f_{n+1}f_{n+2})$ was used in \cite{Vekslerchik} to put evolution equations of the  Volterra lattice hierarchy (\ref{67531}) into bilinear form.
\end{remark}
\begin{remark}
According to Theorem 1.1 from \cite{Hone2},  the identities 
\[
f_nf_{n+2s+1}=-G^{(s)} f_{n+1}f_{n+2s}+K^{(s)} f_{n+s}f_{n+s+1}\;\forall\;\;\mbox{even}\; s\geq 2
\]
and
\[
f_nf_{n+2s+2}=\left(G^{(s)}\right)^2 f_{n+2}f_{n+2s}+K^{(s)} f^2_{n+s+1}\;\forall\;\;\mbox{odd}\; s\geq 3
\]
are valid, where $K^{(s)}$ is the first integral of $s$-th equation (\ref{4}) defined by \cite{Hone2}
\[
K^{(s)}=\left(\prod_{j=0}^{s-1}y_{n+2j}+G^{(s)}\right)\prod_{q=0}^{(s-2)/2} \prod_{j=1}^{s-1}y_{n+2q+j},\;\mbox{if}\;s\;\mbox{even}
\]
and
\[
K^{(s)}=\left(\prod_{j=0}^{2s-1}y_{n+j}-\left(G^{(s)}\right)^2\right)\prod_{q=0}^{(s-3)/2} \prod_{j=2}^{s}y_{n+2q+j},\;\mbox{if}\;s\;\mbox{odd}.
\]
\end{remark}

\begin{remark}
Equation (\ref{9}) represents a slight generalization of the Somos-$4$ equation
\begin{equation}
f_nf_{n+4}=f_{n+1}f_{n+3}+f_{n+2}^2, 
\label{54309}
\end{equation}
that generate, for example,  the sequence of positive integers that begins with $(1, 1, 1, 1, 2, 3, 7, 23, 59,\ldots)$. Despite the fact that, generally speaking, this recurrence relation must generate a sequence of fractional numbers, they all turns out to be integers.
A proof of  integrity of this  sequence  known as the Somos-4 sequence can be found  in \cite{Gale}. It has now become clear that the reason for  integrity of this sequence lies in the fact that equation (\ref{54309}) has the Laurent property. In this connection, it is appropriate to mention here  the relationship of equation (\ref{54309}) to cluster algebras \cite{Fordy}, \cite{Fordy1}.  
One of the properties of  equation (\ref{8}) is that it has also the Laurent property (cf. \cite{Hone2}), that  means that $f_n\in \mathbb{Z}[\alpha_0^{(s)}, \alpha_1^{(s)}, f_0^{\pm 1},\ldots, f_{s+1}^{\pm 1}]$ for any $n\geq s+2$. The Laurent property for equation (\ref{8}), in fact, follows, as a special case, from this property  for the Hirota-Miwa equation \cite{Mase}.
\end{remark}

In the sequel we need the following lemma.
\begin{lemma} \label{8530}  
Any solution $\{w_n\}$ to  equation (\ref{10}) is also a solution of the equation
\begin{equation}
w_{n}w_{n+1}^2w_{n+2}^2w_{n+3}=\alpha_n^{(3)} w_{n+1}w_{n+2}+G^{(3)}
\label{101}
\end{equation}
with
\begin{equation}
\alpha_0^{(3)}=\alpha_1^{(3)}=-G^{(2)}\;\; \mbox{and}\;\; G^{(3)}=G_1^{(2)}=\alpha_0^{(2)}\alpha_1^{(2)}+G^{(2)} H^{(2)}.
\label{111}
\end{equation}
In addition, in these circumstances, we have  $H^{(3)}=H^{(2)}$.
\end{lemma}
\textbf{Proof.} 
Provided that $\{w_n\}$ is any solution of equation (\ref{10}), for some values of $\alpha_0^{(2)}, \alpha_1^{(2)}$ and $G^{(2)}$, we get the following: 
\begin{eqnarray}
&&w_{n}w_{n+1}^2w_{n+2}\cdot w_{n+1}w_{n+2}^2w_{n+3} \nonumber\\
&&\;\;\;\;\;\;\;\;\;\;\;\;\;\;=\left(\alpha_n^{(2)} w_{n+1} + G^{(2)}\right)\left(\alpha_{n+1}^{(2)}  w_{n+2} + G^{(2)}\right) \nonumber\\
&&\;\;\;\;\;\;\;\;\;\;\;\;\;\;=\alpha_n^{(2)}\alpha_{n+1}^{(2)} w_{n+1}w_{n+2}+G^{(2)}\left(\alpha_n^{(2)} w_{n+1}+\alpha_{n+1}^{(2)} w_{n+2}\right)+\left(G^{(2)}\right)^2. \nonumber
\end{eqnarray}
Taking into account (\ref{9809}), we get 
\[
w_{n}w_{n+1}^2w_{n+2}^2w_{n+3} =-G^{(2)} w_{n+1}w_{n+2}+\left(\alpha_n^{(2)}\alpha_{n+1}^{(2)}+G^{(2)} H^{(2)}\right). 
\]
Whence it follows that if $\{w_n\}$ satisfies equation (\ref{10}), then it also satisfies (\ref{101}) with corresponding coefficients given by (\ref{111}). 

Moreover, by  (\ref{10}) and (\ref{761}), we have
\begin{eqnarray*}
H^{(3)}-H^{(2)}&=&w_{n+2}w_{n+3}+\frac{\alpha^{(3)}_{n+1}}{w_{n+1}w_{n+2}}-\frac{\alpha^{(2)}_{n+1}}{w_{n+1}} \\
&=&\left(\frac{\alpha^{(2)}_{n+1}}{w_{n+1}}+\frac{G^{(2)}}{w_{n+1}w_{n+2}}\right)+\frac{\alpha^{(3)}_{n+1}}{w_{n+1}w_{n+2}}-\frac{\alpha^{(2)}_{n+1}}{w_{n+1}}.
\end{eqnarray*}
The latter is zero by (\ref{111}). Therefore this lemma is proved. $\Box$

Let us notice that equation (\ref{101})  is not a simple consequence of equation (\ref{10}), as it may seem at the first glance, since $G^{(3)}$ depends not only on coefficients of equation (\ref{10}) but also on $H^{(2)}$, that is, these coefficients depend on the solution itself. Also, it should be noted that   $\alpha_n^{(3)}$ does not depend on $n$ provided that  $\{w_n\}$ is a solution of  equation (\ref{10}).
\begin{theorem} \label{853098401}  
Any solution $\{w_n\}$ to  equation (\ref{10}) is also a solution to  equation (\ref{7}) for any $s\geq 4$ with suitable  coefficients $\alpha_0^{(s)}, \alpha_1^{(s)}$ and $G^{(s)}$. These coefficients are defined by  the following recurrent relations:
\begin{equation}
\alpha_n^{(s+2)}=\frac{G^{(s+1)}}{G^{(s)}}\alpha_{n}^{(s)}\;\;\mbox{and}\;\;  G^{(s+2)}=G_1^{(s+1)}-\frac{G^{(s+1)}}{G^{(s)}}G_1^{(s)}\;\; s\geq 2,
\label{8941}
\end{equation}
where  $G_1^{(s)}=G^{(s)}H^{(s)}+\alpha_0^{(s)}\alpha_1^{(s)}$.
In addition, the first integrals $H^{(s)}$ are defined by the recurrent relation
\begin{equation}
H^{(s+2)}=H^{(s+1)}+\frac{G^{(s+1)}}{G^{(s)}}\;\;  s\geq 2. 
\label{894111}
\end{equation}
\end{theorem}
The proof of this theorem is quite technical and cumbersome and therefore we carried it to the Appendix.  Evidently recurrent relations (\ref{8941}) and (\ref{894111}) supplemented by (\ref{111}) unambiguously define coefficients $\alpha_{0}^{(s)}, \alpha_{1}^{(s)}, G^{(s)}$, for all $s\geq 4$ as rational functions of $\alpha_{0}^{(2)}, \alpha_{1}^{(2)}, G^{(2)}$ and $H^{(2)}$. 
Moreover, from Lemma \ref{8530}  and Theorem \ref{853098401}, we  get the fact that, provided $\{w_n\}$ is any solution of equation (\ref{10}),  we have $\alpha_0^{(s)}=\alpha_1^{(s)}$  for all odd $s\geq 3$. It is easy to derive the following recurrence relation:
\begin{eqnarray}
G_1^{(s+2)}&=&G_1^{(s+1)}H^{(s+1)}+\frac{G^{(s+1)}}{G^{(s)}}\left(G_1^{(s+1)}-G_1^{(s)}H^{(s+1)}\right) \nonumber \\
&&-\left(\frac{G^{(s+1)}}{G^{(s)}}\right)^2G^{(s)}H^{(s)}\;\; s\geq 2. \label{9944}
\end{eqnarray}

As a consequence of Lemma \ref{8530}  and Theorem \ref{853098401}, we get the following.  
\begin{theorem} \label{64276}
\begin{itemize}
\item[1)]
Any solution $\{y_n\}$ to equation (\ref{109}) is also a solution to  equation (\ref{1555}) for  $s=3$ with $H^{(3)}=H^{(2)}$. The first integrals, in this case, are expressed via $G^{(2)},\;\; G_1^{(2)}$ and $H^{(2)}$ as 
\[
G^{(3)}=G_1^{(2)}\;\; \mbox{and}\;\; G_1^{(3)}=G_1^{(2)}H^{(2)}+\left(G^{(2)}\right)^2,
\]
where 
\[
G^{(2)}=y_1\left(y_0+y_1+y_2-H^{(2)}\right)
\]
and
\[
G_1^{(2)}=y_0y_1y_2+y_1^2\left(y_0+y_1+y_2-H^{(2)}\right);
\]
\item[2)]
any solution $\{y_n\}$ to equation (\ref{109}) is also a solution to  equation (\ref{1555}) for all  $s\geq 4$, where  $H^{(s)}$ is defined by recurrent relation (\ref{894111}) supplemented by (\ref{9944}) and the second relation in (\ref{8941}).
\end{itemize}
\end{theorem}
It is evident that, under the condition of Theorem \ref{64276}, $H^{(s)}$, for all $s\geq 4$, are unambiguously expressed as a rational function of variables $(y_0, y_1, y_2, H^{(2)})$.
\begin{theorem} \label{6427611}
Any solution $\{f_n\}$ to equation (\ref{9}) is also a solution to  equation (\ref{8}) for all  $s\geq 3$, where the coefficients $\alpha_n^{(s)}$ and $G^{(s)}$ are calculated via (\ref{8941}).
\end{theorem}
\textbf{Proof.} We  prove this theorem by a contradiction method. To this aim, we  assume,  that $\{f_n\}$ is a solution to (\ref{9}) but at the same time it is not a solution 
to $s$-th equation (\ref{8}) for any fixed $s\geq 3$. Due to our assumption $\{w_n=f_nf_{n+2}/f_{n+1}^2\}$ is a solution to equation (\ref{10}), and by Theorem \ref{853098401}, it  is also a solution to  $s$-th equation (\ref{7}). But if, according to our assumption, $\{f_n\}$ is not  a solution to $s$-th equation (\ref{8}), then $\{w_n\}$ can not be a solution to $s$-th equation (\ref{7}). Thus, our assumption leads to a contradiction. $\Box$

\section{Construction of solution for difference equations} \label{888}

In this section, we  give an example of a solution for equation (\ref{10}). To this aim, let us define  the sequence $\{T_n : n\geq 0\}$  by linear recurrent relation (\ref{45321})
with arbitrary initial condition, where $P_0, P_1$ and $Q$ are supposed to be arbitrary numbers. It is obvious that  $T_n$ can be written as a linear combination $T_n=T_0N_n+T_1D_n$, where $D_n=D_n(P_0, P_1, Q)$ and  $N_n=N_n(P_0, P_1, Q)$ represent particular solutions of (\ref{45321}) with initial conditions $(0, 1)$ and  $(1, 0)$, respectively.
For example, 
\[
D_0=0,\;\; D_1=1,\;\; D_2=P_0,\;\; D_3=P_0P_1+Q,\;\; D_4=P_0\left(P_0P_1+2Q\right),\ldots
\]
and
\[
N_0=1,\;\; N_1=0,\;\; N_2=Q,\;\; N_3=QP_1,\;\; N_4=Q\left(P_0P_1+Q\right),\ldots
\]
To clarify the meaning of these polynomials,  for any $P_0, P_1$ and $Q$, let us consider the following 2-periodic ``continued fraction'': 
\begin{equation}
C=\frac{Q}{P_0+\displaystyle{\frac{Q}{P_1+\displaystyle{\frac{Q}{P_0+\displaystyle{\frac{Q}{P_1+\cdots }}}}}}}
\label{5432}
\end{equation}
and its  convergents 
\[
C_2=\frac{Q}{P_0},\;\; C_3=\frac{Q}{P_0+\displaystyle\frac{Q}{P_1}},\;\; C_4=\frac{Q}{P_0+\displaystyle\frac{Q}{P_1+\displaystyle\frac{Q}{P_0}}},\ldots
\]
Then, as is known,  $C_n=N_n/D_n\;\; \forall n\geq 2$. Of course, $C$ is literally a continued fraction only if $P_0, P_1$ and $Q$ are integers.

\begin{remark}
For the polynomials $D_n(P_0, P_1, Q)$ and  $N_n(P_0, P_1, Q)$, one knows the following (see, for example, \cite{Bala}). The polynomial $D_{2n}/P_0$  and $D_{2n+1}$, for any $n\geq 1$, are in fact homogeneous polynomial in $P_0P_1$ and $Q$. Moreover, the sequence  $\{D_n : n\geq 0\}$ is a divisibility  sequence. This fact can be illustrated by the formula
\[
D_n=\prod_{d | n}\Phi_d(P_0P_1, Q),
\] 
where, for example,
\[
\Phi_1=1,\;\; \Phi_2=P_0, \;\;\Phi_3=P_0P_1+Q,\;\; \Phi_4=P_0P_1+2Q,
\]
\[
\Phi_5=P_0^2P_1^2+3P_0P_1Q+Q^2,\;\; \Phi_6=P_0P_1+3Q,\ldots 
\]

Let us define the sequence of polynomials $U_n=U_n(L, M)$ by recurrence relations
\begin{equation}
U_{2n}=U_{2n-1}-M U_{2n-2}\;\;\mbox{and}\;\;    U_{2n+1}=L U_{2n}-M U_{2n-1}\;\; n\geq 1
\label{453266666}
\end{equation}
with $U_0=0$ and  $U_1=1$. For example,
\[
U_2=1,\;\; U_3=L-M,\;\; U_4=L-2M,\;\; U_5=L^2-3LM+M^2,\ldots
\]
It is known  \cite{Bala} that 
\[
D_n(P_0, P_1, Q)=\left\{
\begin{array}{l}
U_n(P_0P_1, -Q),\;\;\mbox{if}\;\;n\;\;\mbox{odd}\\[0.1cm]
P_0 U_n(P_0P_1, -Q),\;\;\mbox{if}\;\;n\;\;\mbox{even}
\end{array}
\right.
\]
and
\[
N_{n+1}(P_0, P_1, Q)=\left\{
\begin{array}{l}
QU_n(P_0P_1, -Q),\;\;\mbox{if}\;\;n\;\;\mbox{odd}\\[0.1cm]
QP_1U_n(P_0P_1, -Q),\;\;\mbox{if}\;\;n\;\;\mbox{even},
\end{array}
\right.
\]
and consequently $N_{n+1}(P_0, P_1, Q)=QD_n(P_1, P_0, Q)$. 
\end{remark}

To prepare the ground  to prove Theorem \ref{4387} below, let us formulate some lemmas.
\begin{lemma} By (\ref{45321}), the polynomials $T_n$ satisfy the identity
\begin{equation}
T_{2n+2s}=D_{2s+2}T_{2n-1}+QD_{2s+1}T_{2n-2}
\label{776}
\end{equation}
and
\begin{equation}
T_{2n+2s+1}=\frac{P_1}{P_0}D_{2s+2}T_{2n}+QD_{2s+1}T_{2n-1}
\label{7761}
\end{equation}
for all $s\geq 0$.
\end{lemma}
\textbf{Proof.} Let us prove identity (\ref{776}) by induction on $s$. To this aim, let us suppose that relations (\ref{776}) and (\ref{7761}) is valid for some $s\geq 0$.  By the first relation in (\ref{45321}), we have
\begin{eqnarray*}
T_{2n+2s+2}&=&P_0T_{2n+2s+1}+QT_{2n+2s}\\
&=&P_0\left(\frac{P_1}{P_0}D_{2s+2}T_{2n}+QD_{2s+1}T_{2n-1}\right)\\
&&+Q\left(D_{2s+2} T_{2n-1}+QD_{2s+1}T_{2n-2}\right)\\
&=&P_1D_{2s+2} T_{2n}+Q\left(P_0D_{2s+1}+D_{2s+2}\right)T_{2n-1}+Q^2D_{2s+1}T_{2n-2}\\
&=&P_1D_{2s+2}\left(QT_{2n-2}+P_0T_{2n-1}\right)+Q\left(P_0D_{2s+1}+D_{2s+2}\right)T_{2n-1}\\
&&+Q^2D_{2s+1}T_{2n-2}\\
&=&\left(P_0\left(P_1D_{2s+2}+QD_{2s+1}\right) +QD_{2s+2}\right)T_{2n-1}\\
&&+Q\left(P_1D_{2s+2}+QD_{2s+1}\right)T_{2n-2}\\
&=&D_{2s+4}T_{2n-1}+QD_{2s+3}T_{2n-2}.
\end{eqnarray*}
Thus, we have proved that if  relations (\ref{776}) and (\ref{7761}) are valid for some $s\geq 0$, then (\ref{776}) is also valid for $s+1$.
Relations (\ref{776}) and (\ref{7761}), for $s=0$, are evidently valid by the condition of the lemma. Thus, by induction on $s$, we prove this lemma. Identity (\ref{7761}) can be proved  analogously.  $\Box$ 

We actually need the following consequence of this lemma. Since $\{D_n\}$ represents a particular solution of linear equations (\ref{45321}), then  from (\ref{776}), as a particular case, we derive the following relation: 
\begin{equation}
D_{2n+2s}=D_{2s+2}D_{2n-1}+QD_{2s+1}D_{2n-2}
\label{6754}
\end{equation}
for all $s\geq 0$ and $n\geq 1$. As a consequence of (\ref{6754}) we get the following lemma.
\begin{lemma} The polynomials $D_r(P_0, P_1, Q)$ satisfy the identity
\begin{equation}
D_{r}\left(D_{r+3}-Q D_{r+1}\right)=D_{r+2}\left(D_{r+1}-Q D_{r-1}\right)\; \forall r\geq 1.
\label{7761223}
\end{equation}
\end{lemma} 
\textbf{Proof.}  Indeed, assuming $n=k+2, s=k-1$ and $n=k+1, s=k$  in (\ref{6754}),  we derive the identities 
\begin{eqnarray*}
D_{4k+2}&=&D_{2k}D_{2k+3}+QD_{2k-1}D_{2k+2}\\
&=&D_{2k+1}D_{2k+2}+QD_{2k}D_{2k+1}
\end{eqnarray*}
that gives (\ref{7761223}) for $r=2k$. In turn, assuming $n=s=k+1$ and $n=k+2, s=k$ with some $k\geq 1$ in (\ref{6754}),  we derive the identities 
\begin{eqnarray*}
D_{4k+4}&=&D_{2k+1}D_{2k+4}+QD_{2k}D_{2k+3}\\
&=&D_{2k+2}D_{2k+3}+QD_{2k+1}D_{2k+2}
\end{eqnarray*}
that gives (\ref{7761223}) for $r=2k+1$.
$\Box$

The following lemma plays crucial role for constructing some class of solutions for difference equation (\ref{7}).
\begin{lemma} \label{672}
A  sequence $\{T_n : n\geq 0\}$ determined by the recursion (\ref{45321}) is also a solution of bilinear equation (\ref{9}), provided that
\begin{equation}
\alpha^{(2)}_n=-\frac{P_n^2}{Q}\;\;\mbox{and}\;\; G^{(2)}=\frac{P_0P_1+Q}{Q}.
\label{7223}
\end{equation}
Moreover, we have 
\begin{equation}
H^{(2)}=\frac{-P_0P_1+2Q}{Q}.
\label{72231}
\end{equation}
\end{lemma} 
\textbf{Proof.} 
This lemma can be proved by direct but quite cumbersome calculations. $\Box$ 

It should be noted that the parameters $\alpha^{(2)}_0, \alpha^{(2)}_1$ and $G^{(2)}$ in (\ref{7223})  do not depend on $T_0$ and $T_1$. 

As a consequence of Lemma \ref{672}, we derive the following two identities: 
\begin{equation}
P_0^2D_{s}D_{s+2}=\left(P_0P_1+Q\right)D_{s+1}^2-QD_{s-1}D_{s+3}\;\;\forall\;\mbox{odd}\;\; s\geq 1
\label{731}
\end{equation}
and
\begin{equation}
P_1^2D_{s}D_{s+2}=\left(P_0P_1+Q\right)D_{s+1}^2-QD_{s-1}D_{s+3}\;\;\forall\;\mbox{even}\;\; s\geq 2.
\label{73123}
\end{equation}

\begin{lemma} \label{5643} 
The relations
\begin{eqnarray}
&&Q  D_s D_{s+1}\left(D_{s+1}D_{s+4}-D_{s+2} D_{s+3}\right)  + D_{s+2}D_{s+3}\left(D_s D_{s+3} -D_{s+1} D_{s+2}\right) \nonumber \\
&&\;\;\;\;\;\;\;\;\;\;\;\;\;\;+P_0^2D_{s+1}^3 D_{s+3}-P_1^2 D_s D_{s+2}^3 =0\;\; \forall\; \mbox{even}\; s\geq 2 
\label{409654}
\end{eqnarray}
and
\begin{eqnarray}
&&Q  D_s D_{s+1}\left(D_{s+1}D_{s+4}-D_{s+2} D_{s+3}\right)  + D_{s+2}D_{s+3}\left(D_s D_{s+3} -D_{s+1} D_{s+2}\right) \nonumber \\
&&\;\;\;\;\;\;\;\;\;\;\;\;\;\;+P_1^2D_{s+1}^3 D_{s+3}-P_0^2 D_s D_{s+2}^3 =0\;\; \forall\; \mbox{odd}\; s\geq 1 
\label{409654111}
\end{eqnarray}
are identities.
\end{lemma}
\textbf{Proof.} Let us prove only (\ref{409654111}). By (\ref{731}), for all odd $s\geq 1$, we have 
\begin{eqnarray*}
P_1^2D_{s+1}^3 D_{s+3}-P_0^2 D_s D_{s+2}^3&=&D_{s+1}^2\left\{(P_0P_1+Q)D_{s+2}^2-QD_{s}D_{s+4}\right\} \\
&&-D_{s+2}^2\left\{(P_0P_1+Q)D_{s+1}^2-QD_{s-1}D_{s+3}\right\} \\
&=&Q\left(D_{s-1} D_{s+2}^2D_{s+3}-D_{s} D_{s+1}^2D_{s+4}\right).
\end{eqnarray*}
Taking the latter into account, we can rewrite  (\ref{409654111}) as 
\begin{eqnarray*}
&&Q  D_s D_{s+1}\left(D_{s+1}D_{s+4}-D_{s+2} D_{s+3}\right)  + D_{s+2}D_{s+3}\left(D_s D_{s+3} -D_{s+1} D_{s+2}\right) \\
&&\;\;\;\;\;\;\;\;\;\;\;\;\;\;+Q\left(D_{s-1}D_{s+2}^2D_{s+3}-D_{s}D_{s+1}^2D_{s+4}\right) =0
\end{eqnarray*}
or quite simply as $QJ_{s-1}+J_s=0$, where, by definition,  $J_s=D_{s}D_{s+3}-D_{s+1}D_{s+2}$. But the latter is nothing but identity   (\ref{7761223}), which has already been proven. Identity (\ref{409654}) can be proved  using similar reasoning.
$\Box$

It follows, from  Lemma \ref{672},  that the sequence $\{w_n=T_nT_{n+2}/T_{n+1}^2 : n\geq 0\}$  represents  a particular solution of equation (\ref{10}).
In turn, by Theorem  \ref{853098401}, we know that this sequence is also a solution of equation (\ref{7}) for all $s\geq 3$. More exactly, we have the following theorem.
\begin{theorem} \label{4387} 
The sequence $\{w_n=T_nT_{n+2}/T_{n+1}^2 : n\geq 0\}$ is a particular solution of equation (\ref{7}), for any $s\geq 2$, provided that the parameters $\alpha_0^{(s)}, \alpha_1^{(s)}$ and $G^{(s)}$ are given by
\begin{equation}
\alpha_0^{(s)}=-\frac{P_0D_{s}}{Q D_{s-1}},\;\;
\alpha_1^{(s)}=\left\{
\begin{array}{l}
\alpha_0^{(s)},\;\;\mbox{if}\; $s$\; \mbox{odd}\\[0.2cm] \displaystyle
\frac{P_1^2}{P_0^2}\alpha_0^{(s)},\;\;\mbox{if}\; $s$\; \mbox{even}
\end{array}
\right. \;\;\mbox{and}\;\; G^{(s)}=\frac{D_{s+1}}{Q D_{s-1}}. 
\label{723}
\end{equation}
In addition we have 
\begin{equation}
H^{(s)}=-G^{(s)}+s+1.
\label{5632}
\end{equation}
\end{theorem}
\textbf{Proof.} To begin with, we note that formulas (\ref{723}) and (\ref{5632}), in the case $s=2$, give what we have in (\ref{7223}) and  (\ref{72231}). Moreover, by (\ref{111}), we get
\[
\alpha_0^{(3)}=\alpha_1^{(3)}=-\frac{P_0P_1+Q}{Q}\;\; \mbox{and}\;\; G^{(3)}=\frac{P_0P_1+2Q}{Q}
\]
and this  corresponds to  (\ref{723}) for $s=3$. Also, one can check that the equality  $H^{(3)}=H^{(2)}$ is in accordance to  (\ref{5632}). It remains  to check that the substitution of (\ref{723}) and (\ref{5632}) into recurrent relations  (\ref{8941}) and (\ref{894111}) in Theorem \ref{853098401} gives  identities. 

It is easy to check that substituting $\alpha_0^{(s)}$ and $\alpha_1^{(s)}$ given by (\ref{723}) into (\ref{8941}) gives trivial identities. 
In turn, substituting  (\ref{5632})  into (\ref{894111})  yields the relation
\[
G^{(s+1)}=G^{(s)}-\frac{G^{(s)}}{G^{(s-1)}}+1.
\]
Expressing here $G^{(s)}$ via $D_s$ as in (\ref{723}), we get the identity (\ref{7761223}) that has already been  proven. So it remains  to check the fulfillment of the second recurrent relation     in (\ref{8941}). One can make sure that substituting (\ref{723}) and (\ref{5632})   into this relation  gives  identities (\ref{409654}) and (\ref{409654111}) that has already proven. $\Box$

In turn, Theorem \ref{4387} implies the following theorem.
\begin{theorem} \label{890654}
The  sequence $\{y_n=T_nT_{n+3}/(T_{n+1}T_{n+2}) : n\geq 0\}$ is a solution of the equation (\ref{1555}) with parameter $H^{(s)}$ given by (\ref{5632}) for all $s\geq 2$.
\end{theorem} 
The following theorem characterizes the  solution provided by the Theorem \ref{890654}.
\begin{theorem} 
The  sequence $\{y_n=T_nT_{n+3}/(T_{n+1}T_{n+2}) : n\geq 0\}$ satisfies the relation 
\begin{equation}
y_n+y_{n+1}+y_{n+2}-y_ny_{n+2}+\frac{1}{y_{n+1}}=3.
\label{765098}
\end{equation}
\end{theorem} 
\textbf{Proof.} We have
\begin{equation}
G_1^{(2)}=G^{(3)}=\frac{P_0P_1+2Q}{Q}=y_ny_{n+1}y_{n+2}+y_{n+1}^2\left(y_n+y_{n+1}+y_{n+2}-H^{(2)}\right).
\label{65444234}
\end{equation} 
By the relation
\begin{equation}
G^{(2)}=\frac{P_0P_1+Q}{Q}=y_{n+1}\left(y_n+y_{n+1}+y_{n+2}-H^{(2)}\right),
\label{65444}
\end{equation} 
we can rewrite (\ref{65444234}) as 
\[
\frac{P_0P_1+2Q}{Q}=y_ny_{n+1}y_{n+2}+y_{n+1}\frac{P_0P_1+Q}{Q}.
\]
In turn we can rewrite this relation as
\[
y_{n+1}\left(y_ny_{n+2}-H^{(2)}-\frac{1}{y_{n+1}}+3\right)=\frac{P_0P_1+Q}{Q}=G^{(2)}
\]
Finally comparing the latter with  (\ref{65444}), we get (\ref{765098}). $\Box$

Of course, as a particular case we get the following relation:
\[
y_0+y_1+y_2-y_0y_2+\frac{1}{y_1}=3
\]
that allows to parametrize the solution in terms of initial values.

At the end of this section, we would like to give  examples of the sequence $\{T_n : n\geq 0\}$. Let, for example,  $P_0=P_1=Q=1$.  Then recurrent relations (\ref{45321}) is reduced to the relation $T_{n+2}=T_{n+1}+T_n$, that defines  generalized Fibonacci numbers. In particular, if $T_0=0$ and $T_1=1$, then $T_n=D_n=F_n$, where $F_n$'s are the Fibonacci numbers. In turn, if $T_0=2$ and $T_1=1$, then $T_n=2N_n+D_n=L_n$ , where $L_n$'s are Lucas numbers. Evidently, that $L_n=F_n+2F_{n-1}$.

Continued fraction (\ref{5432}), in this case,  becomes  
\[
C=\Phi-1=\frac{1}{1+\displaystyle{\frac{1}{1+\displaystyle{\frac{1}{1+\cdots}}}}},
\]
where ${\displaystyle \Phi ={\frac {{\sqrt {5}}+1}{2}}}$ is well known as a golden section. The sequence of convergents of $C$, in this case, are expressed via the Fibonacci numbers as $C_n=F_{n-1}/F_n$. 

More generally, let $P_0=P_1=P$. Let $P$ and $Q$ be relatively prime integers such that discriminant $D=P^2+4Q\neq 0$ of the characteristic equation $x^2-Px-Q=0$ is not zero. Let $a$ and $b$ be algebraic numbers such that $a+b=P$ and $ab=-Q$. Finally, if a ratio $a/b$ is not a root of unity then a Lucas sequence and its  companion  are defined as \cite{Lehmer}
\[
D_n=\frac{a^n-b^n}{a-b}\;\; \mbox{and}\;\; 2N_n+P D_n=a^n+b^n,
\]
respectively.

Also as a special case, the sequence  $\{T_n : n\geq 0\}$ contains a Lehmer sequence, which, to a certain extent, has already been mentioned above. Let $P_0=1$. Let us denote $P_1=L$ and $Q=-M$. Then 
linear recurrent relations (\ref{45321}) reduce to relations (\ref{453266666}). 
Let $L$ and $M$ be relatively prime integers such that $D=L-4M\neq 0$, while  $a$ and $b$ be algebraic numbers that are roots of the characteristic equation $x^2-\sqrt L x+M=0$, that is, $L=(a+b)^2$ and $M=ab$. In addition, let us suppose that the quotient  $a/b$ is not a root of unity,  then the Lehmer sequence   is defined as \cite{Lehmer}
\[
D_n=\left\{
\begin{array}{c}
\displaystyle
\frac{a^n-b^n}{a-b},\;\;\mbox{if}\;n\;\mbox{odd} \\[0.3cm]
\displaystyle
\frac{a^n-b^n}{a^2-b^2}, \;\;\mbox{if}\;n\;\mbox{even}.
\end{array}
\right. 
\]
To define a companion Lehmer sequence one needs to put $P_0=L$, $P_1=1$ and $Q=-M$. Then
\[
2N_n+D_n=\left\{
\begin{array}{c}
\displaystyle
\frac{a^n+b^n}{a+b},\;\;\mbox{if}\;n\;\mbox{odd} \\[0.3cm]
\displaystyle
a^n+b^n, \;\;\mbox{if}\;n\;\mbox{even}.
\end{array}
\right. 
\]

\section{Final remarks}


Let us briefly discuss the results presented in this paper. In the end, we find the relationship of integer sequences with integrable equations (\ref{1555}). Such relationship can later find its applications in the theory of integer sequences. 
It would also be interesting to explore these results for investigation Volterra lattice integrable hierarchy (\ref{67531}).

Difference equations (\ref{1555}) represent, in fact, a subclass of two-parametric class of equations \cite{Svinin1}
\begin{equation}
y_{n+k}\tilde{S}^k_{s+1}(n)=y_{n+s}\tilde{S}^k_{s+1}(n+1)\;\; k\geq 1,\; s\geq k+1.
\label{49867}
\end{equation}
A precise description of these equations is as follows. Discrete polynomial $\tilde{S}^k_s(n)$ in (\ref{49867}) can be written as 
\[
\tilde{S}^k_s(n)=\sum_{j=0}^k(-1)^jH_j^{(k, s)}S^{k-j}_{s-j}(n+j),
\]
where $H_j^{(k, s)}$ are some parameters and  $H_0^{(k, s)}=1$, while  discrete polynomials  $S^k_s(n)$  are defined by (\ref{6753}). Evidently the equations  (\ref{1555}) can be rewritten as
\[
y_{n+1}\tilde{S}^1_{s+1}(n)=y_{n+s}\tilde{S}^1_{s+1}(n+1),\; s\geq 2.
\]

Actual calculations suggest that Theorem \ref{64276} can be expanded to this class of equations. More exactly, we suppose that any solution to equation (\ref{109}) is at the same time a solution to equation (\ref{49867}), for all $k\geq 1$ and $s\geq k+1$, provided that we restrict these parameters  appropriately. Moreover we suppose that this  constraint looks like that
\[
H_k^{(k, s)}=\sum_{j=1}^{k-1} a_j^{(k, s)} H_j^{(k, s)}+b^{(k, s)},
\]
where the coefficients $a_j^{(k, s)}$ and $b^{(k, s)}$ are some rational functions of variables $(y_0, y_1, y_2, H^{(2)})$.
Now we do not see a direct way to make such a generalization, but, in the future, we intend to continue our investigation in this direction.

\appendix

\section{}

Let us prove here Theorem \ref{853098401}, by induction. For any $s\geq 4$, by (\ref{7}), we have
\begin{eqnarray*} 
&&\prod_{j=0}^{s-1} w_{n+j}w_{n+j+1} \cdot \prod_{j=0}^{s-1} w_{n+j+1}w_{n+j+2}\\
&&\;\;\;=\left(\alpha_n^{(s)} \prod_{j=1}^{s-1} w_{n+j} + G^{(s)}\right)\left(\alpha_{n+1}^{(s)}  \prod_{j=1}^{s-1} w_{n+j+1}+G^{(s)}\right) \\
&&\;\;\;=\alpha_n^{(s)}\alpha_{n+1}^{(s)} \prod_{j=1}^{s-1} w_{n+j}w_{n+j+1}+G^{(s)}\prod_{j=2}^{s-1} w_{n+j}\left(\alpha^{(s)}_n w_{n+1}+\alpha^{(s)}_{n+1} w_{n+s}\right) \\
&&\;\;\;\;\;\;+\left(G^{(s)}\right)^2.  
\end{eqnarray*}
We can rewrite (\ref{765}) as
\begin{eqnarray*} 
\alpha^{(s)}_n w_{n+1}+\alpha^{(s)}_{n+1} w_{n+s}&=&H^{(s)}\prod_{j=1}^{s} w_{n+j}-\frac{G^{(s)}}{\prod_{j=2}^{s-1} w_{n+j}}\\
&&-\prod_{j=1}^{s} w_{n+j}\sum_{j=1}^{s-1} w_{n+j}w_{n+j+1}.
\end{eqnarray*}
Taking the latter into account, we get 
\begin{eqnarray*} 
&&\prod_{j=0}^{s-1} w_{n+j}w_{n+j+1} \cdot \prod_{j=0}^{s-1} w_{n+j+1}w_{n+j+2}\\
&&\;\;\;\;\;\;\;\;\;\;\;\;=\alpha_n^{(s)}\alpha_{n+1}^{(s)} \prod_{j=1}^{s-1} w_{n+j}w_{n+j+1}+G^{(s)}\prod_{j=2}^{s-1} w_{n+j}\left(H^{(s)}\prod_{j=1}^{s} w_{n+j}\right.\\
&&\;\;\;\;\;\;\;\;\;\;\;\;\;\;\;\left. -\frac{G^{(s)}}{\prod_{j=2}^{s-1} w_{n+j}} -\prod_{j=1}^{s} w_{n+j}\sum_{j=1}^{s-1} w_{n+j}w_{n+j+1}\right)+\left(G^{(s)}\right)^2.
\end{eqnarray*}
From this follows the fact that if $\{w_n\}$ satisfies equation (\ref{7}), for some $s\geq 4$, then it satisfies also the equation
\begin{equation}
\prod_{j=0}^{s} w_{n+j}w_{n+j+1} = \left(\alpha_n^{(s)}\alpha_{n+1}^{(s)}+G^{(s)}H^{(s)}\right)-G^{(s)}\sum_{j=1}^{s-1} w_{n+j}w_{n+j+1}. 
\label{943}
\end{equation}
Let us now assume that $\{w_n\}$ is also a solution to $(s-1)$-th equation (\ref{7}). From (\ref{7}) and (\ref{761}), we derive the following relation:
\begin{equation}
\sum_{j=1}^{s-1} w_{n+j}w_{n+j+1}=H^{(s-1)}+\frac{\alpha_{n}^{(s-1)}}{G^{(s-1)}}\left(\alpha_{n+1}^{(s-1)}-\prod_{j=1}^{s} w_{n+j}\right).
\label{7842}
\end{equation}
Substituting (\ref{7842}) into (\ref{943}),  yields
\begin{eqnarray*}
\prod_{j=0}^{s} w_{n+j}w_{n+j+1}&=& \left(\alpha_n^{(s)}\alpha_{n+1}^{(s)}+G^{(s)}H^{(s)}\right)\\
&&-G^{(s)}\left(H^{(s-1)}+\frac{\alpha_{n}^{(s-1)}}{G^{(s-1)}}\left(\alpha_{n+1}^{(s-1)}-\prod_{j=1}^{s} w_{n+j}\right)\right) \\
&=&\alpha_n^{(s+1)} \prod_{j=1}^{s} w_{n+j} + G^{(s+1)},
\end{eqnarray*}
where
\begin{equation}
\alpha_n^{(s+1)}=\frac{G^{(s)}}{G^{(s-1)}}\alpha_{n}^{(s-1)}
\label{6590}
\end{equation}
and
\[
G^{(s+1)}=\alpha_n^{(s)}\alpha_{n+1}^{(s)}+G^{(s)}H^{(s)}-\frac{G^{(s)}}{G^{(s-1)}}\left(\alpha_{n}^{(s-1)}\alpha_{n+1}^{(s-1)}+G^{(s-1)}H^{(s-1)}\right).
\]

In a result, we get the following fact. If $\{w_n\}$  is a solution to $(s-1)$-th and $s$-th equation  (\ref{7}) simultaneously, then it is a solution to  $(s+1)$-th equation. 
On the other hand, from lemma \ref{8530} we  know that if $\{w_n\}$ is a solution of equation  (\ref{10})  then it is also a solution of equations (\ref{101}). It makes  a base of induction.  
Thus, we are in a position  to prove the main part of Theorem \ref{853098401}  by induction on $s$. 

Let us now prove relation (\ref{894111}). Further we use  already proven fact that under the condition of Theorem \ref{853098401}, the sequence $\{w_n\}$ is a solution to the equation (\ref{7}), for any $s\geq 3$,  with  corresponding coefficients. Then, we have
\[
H^{(s+1)}-H^{(s)}=w_{n+s}w_{n+s+1}+\frac{\alpha_{n+1}^{(s+1)}}{\prod_{j=1}^s w_{n+j}}-\frac{\alpha_{n+1}^{(s)}}{\prod_{j=1}^{s-1} w_{n+j}}.
\]  
By (\ref{7}),  we can write
\begin{equation}
w_{n+s}w_{n+s+1}=\frac{\alpha_{n+1}^{(s)}}{\prod_{j=1}^{s-1} w_{n+j}}+\frac{G^{(s)}}{\prod_{j=1}^{s-1} w_{n+j}w_{n+j+1}}.
\label{65901}
\end{equation}
Taking  into account relations (\ref{6590}) and (\ref{65901}), we have
\[
H^{(s+1)}-H^{(s)}=\frac{G^{(s)}}{\prod_{j=1}^{s-1} w_{n+j}w_{n+j+1}}\left(1+\frac{\alpha_{n+1}^{(s-1)}}{G^{(s-1)}}\prod_{j=2}^{s-1} w_{n+j}\right).
\]
Again, by (\ref{7}), 
\[
1+\frac{\alpha_{n+1}^{(s-1)}}{G^{(s-1)}}\prod_{j=2}^{s-1} w_{n+j}=\frac{\prod_{j=1}^{s-1} w_{n+j}w_{n+j+1}}{G^{(s-1)}}
\]
and using the latter we get (\ref{894111}). $\Box$

\section*{Acknowledgments}

I wish to thank the referees for carefully reading the article and for remarks which improved the presentation.

The results were obtained within the framework of the state assignment of the Ministry of Education and Science of the Russian Federation on the project No. 121041300058-1.

\end{document}